# Prediction of Superionic State in LiH$_2$ at Conditions Enroute to Nuclear Fusion

Fude Li (李福德)[1], Hao Wang (王豪)[1], Jinlong Li (李津龙)[1], and Hua Y. Geng (耿华运)[1,2,†]

[1]*National Key Laboratory of Shock Wave and Detonation Physics, Institute of Fluid Physics, CAEP; P. O.Box 919-102 Mianyang, Sichuan, 621900, China*

[2]*HEDPS, Center for Applied Physics and Technology, and College of Engineering, Peking University, Beijing 100871, China*

Hydrogen and lithium, along with their compounds, are crucial materials for nuclear fusion research. High-pressure studies have revealed intricate structural transitions in all these materials. However, research on lithium hydrides beyond LiH has mostly focused on the low-temperature regime. Here, we use density functional theory and *ab initio* molecular dynamics simulations to investigate the behavior of LiH$_2$, a hydrogen-rich compound, near its melting point. Our study is particularly relevant to the low-pressure region of the compression pathway of lithium hydrides toward fusion. We discovered a premelting superionic phase transition in LiH$_2$ that has significant implications for its mass transportation, elastic properties, and sound velocity. The theoretical boundary for the superionic transition and melting temperature was then determined. In contrast, we also found that the primary compound of lithium hydrides, LiH, does not exhibit a superionic transition. These findings have important implications for optimizing the compression path to achieve the ignition condition in inertial confinement fusion research, especially when lithium tritium-deuteride(LiTD) are used as the fuel.

† Corresponding author. E-mail: s102genghy@caep.cn

Chinese Physics B **32**, 106103 (2023)

## 1. Introduction

In inertial confinement fusion (ICF), deuterium- tritium ice or lithium deuteride solid are the most commonly used fuel materials[1][2]. Both contain isotopes of hydrogen and undergo shock or quasi-isentropic loading to achieve the ignition condition[3]. However, during the dynamic compression process, the pressure and temperature increase simultaneously and can intersect the melting curve of the solid fuel[4]. This can have an impact on the ignition, as the quasi-isentropic compression is path-dependent and the structural transition and property change in lithium hydrides along this path can affect subsequent thermodynamics. Furthermore, it is important to note that during the nuclear process, lithium can mutate into polyhydrides such as $LiH_n$[5]. Specifically, the existence of $LiH_2$ at low pressure implies the possibility of using lithium tritium-deuteride(LiTD) as the fuel directly, rather than the tritium-substituted lithium deuteride[Li(DT)], to improve the fuel efficiency. Understanding the structure and behavior of lithium polyhydrides under high pressure and high temperature conditions is crucial for achieving controllable ICF[6].

The structure of $LiH_2$ was first reported by Zurek[7] using density function theory(DFT)[8][9], and the stable pressure is predicted to be 130 to 300 GPa. Recent more careful calculations suggest that $LiH_2$ can be stabilized as a metastable phase down to pressures as low as 10 GPa[10]. In experiment, Howie *et al*.[11] attempted to synthesize polyhydrides by heating LiH mixed with hydrogen up to 600 K and pressurizing to 5 GPa. However, no signal of polyhydrides was observed. Lithium polyhydries were finally synthesized at a condition of 5 GPa and 1800 K in 2015[12][13][14], in which $LiH_2$ was considered as one of the products. The optical spectra measurements of these compounds indicate an insulating nature, which sharply contradicts the metallic state predicted by theoretical methods[7]. This apparent discrepancy was later resolved by using more accurate exchange-correlation functional, which unequivocally demonstrates that it is the lack of appropriate van der Waals interaction[5] that leads to wrong metallic state in these polyhydrides. It also shows that the van der Waals interaction modifies the relative stability order greatly, and two new super-hydrides, $LiH_9$ and $LiH_{10}$, were predicted for the first time[5].

On the other hand, it has been well-established that hydrogen will become metallic when at high pressure beyond 500 GPa[15][16][17]. Doping with metallic elements to form polyhyrides sometimes will lower the metallization pressure greatly, by a mechanism of "chemical pre-compression"[18]. Lithium polyhydrides, which was the first system that was employed to check the concept of chemical pre-compression, is not as promising as other polyhydrides in terms of metallization, mainly due to the strong van der Waals interaction persists in this system, which open the energy gap and prevent the metallization in some stoichiometric compositions[5].

These results exhibit that lithium polyhydrides represent a different type of H-rich compounds other than those of easy to achieve metallization like in lanthanum hydrides[19]. Although "chemical pre-compression" promotes the dissociation of $H_2$ dimers, it is not easy to transition lithium polyhydrides into a metallic state. In this sense, to explore their structural characteristics, it may be helpful in designing low-pressure metallic polyhydrides and searching for superconductors under ambient conditions[15]. Within lithium polyhydrides, $LiH_2$ deserves special attention, not only because its structure is simpler, but also because half of the hydrogen atoms have been dissociated, making it an ideal $H_2+H$ mixture. Its behavior at high temperature could be instructive for understanding other polyhydrides that might be involved in hydrogen storage[20] or ICF.

In this work, we employ *ab initio* molecular dynamics (AIMD) simulations with density functional theory to investigate the structural change, thermodynamic properties and elasticity of $LiH_2$ at high pressure and high temperature. A premelting superionic state is predicted spanning over the whole studied pressure range(from 25 to 300 GPa), as well as the associated elastic softening, all evidence of superionic as shown in Fig.S1. The paper is organized as follows. The computational methods and details are given in Sec. 2. In Sec. 3, the results about the melting point, fractional change of $H_2$ dimer, diffusion coefficient, and elastic constants of $LiH_2$ are discussed, and Sec. 4 summarizes the main findings and the conclusion of the paper.

2. **Computational details**

The structure of LiH$_2$ was taken as that reported by Chen *et al*.[5] which has a symmetry of space group *P4/mbm*, as shown in the inset of Fig. 1 and lattice constants at high pressures are listed in Table S1. The NPT (N-number of particles, P-pressure, and T-temperature) AIMD simulations with Langevin isobaric-isothermal ensemble[21][22][23] were used to equilibrate the structure and lattice constants at different temperatures. The friction coefficient for both Li and H atoms was set to $\gamma = 12 \text{ ps}^{-1}$. A $2\times2\times2$ supercell containing 96 atoms with periodic boundary conditions are employed. In DFT calculations, the cutoff energy of the plane-wave basis was set to 700 eV. The van der Waals corrected PBE exchange-correlation functional of generalized gradient approximation[24][25] as implemented in Vienna Ab initio Simulation Package(VASP)[26] were employed. The k-points were sampled in the first Brillouin zone with a size of 1×1×2 for the Monkhorst-Pack[27] mesh. The AIMD simulations for the structure equilibrating at finite temperature were carried out at the given pressure. Each simulation ran for 10000 steps, with a time step of 0.25 fs.

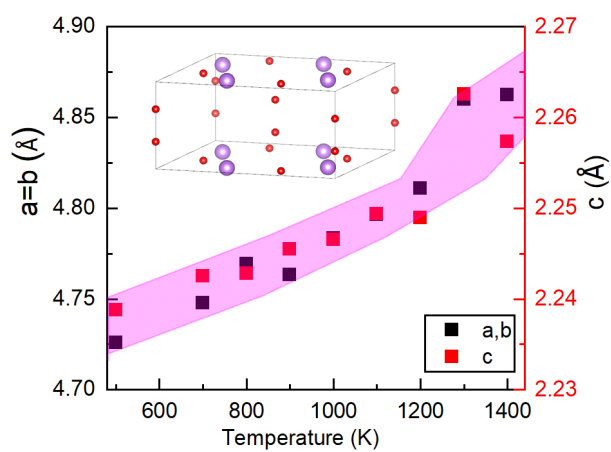

Fig. 1. (Color online) Variation of lattice constants of LiH$_2$ with temperature at 50 GPa, the color-shaded region is guide for the eye. Inset: The conventional unit cell of LiH$_2$ that contains 4 formula units where red balls represent hydrogen atoms and purple balls represent lithium atoms. [Note: the slight fluctuation in NPT simulated lattice constants is due to the small box size and short AIMD simulation we used. However, the over accuracy is better than 1%, and is enough for our purpose in this work.]

After obtained the average lattice constants from the NPT simulation, we utilized a canonical ensemble (NVT) AIMD simulation with a Nose-Hoover thermostat[28][29][30] to calculate the mean squared displacement (MSD) of each type of atoms. The k-points mesh was increased to $2\times2\times3$ to improve the numerical precision. Each NVT simulation also ran 10000 steps, with a time step of 0.25 fs.

In elastic property calculation, we employed long NVT ensemble simulation up to 20000 steps (corresponding to a time duration of 5 ps, with a timestep of 0.25 fs). For each simulation, the first 2000 steps were used to equilibrium the system, and the last 18000 steps were then sampled to compute the statistically averaged stress-strain relationship, from which the elastic constants $C_{ij}$ were derived. In our calculation, the magnitude of strain distortion was set as 0.04. The whole calculation was carried out by using the MyElas[31] code that is designed to compute the finite temperature elasticity. For a tetragonal system as $LiH_2$, six nonequivalent elastic constants $C_{11}$, $C_{12}$, $C_{13}$, $C_{33}$, $C_{44}$, and $C_{66}$ were obtained.

The melting points of $LiH_2$ were determined by using the Z method[32], which has been shown can provide reliable results for anharmonic systems[33][34]. Considering the superionic nature of $LiH_2$ prior to melting, it is reasonable to assume that the system is anharmonic. A large supercell containing 224 H atoms and 112 Li atoms was employed, with each NVE AIMD simulation runs up to 3000 steps (with a timestep of 0.5 fs), which is long enough for the system to equilibrate and achieve ergodicity, as shown in Fig. S2 of the Supplementary Material.

## 3. Results and discussion

### 3.1. Superionic transition

From our AIMD simulations, we found that across the whole pressure range we investigated, the $LiH_2$ always undergoes a superionic transition prior to melting. Only the situation at 50 GPa is illustrated in details here for brevity. As shown in Fig.1, all the lattice vectors increase almost linearly with the increase in temperature from 500 to 1200 K, and then experience a rapid dilation when approaching the melting temperature. In the following simulations, the initial lattice parameters were set to the finite temperature value optimized by NPT simulations. As depicted in Fig.2, below 800 K, both lithium and hydrogen atoms exhibit vibrational motion around their respective equilibrium lattice sites. As the temperature increases further, hydrogen atoms begin to hop between their nearest neighboring sites, while the lithium atoms remain localized at their equilibrium lattice positions. Upon surpassing a temperature of 1100 K, the hydrogen sublattice undergoes complete melting, as evidenced by the MSD displayed in Fig. 2(a). In contrast, the Li sublattice still keeps the solid state under this condition (Fig.2(d)). This is a hall-mark of superionic transition. It should be noted that this transition in $LiH_2$ is a smooth and continuous process that spans over a finite range of 900 – 1200 K. To simplify the discussion, we designate the superionic transition temperature at 50 GPa and 900 K when the H atoms start hopping. This allows clearer communication without any potential ambiguity. The same rule applies to other pressures. Inspecting the AIMD trajectory also indicates that in the early stage of this superionic transition, the H sublattice possesses an 1D chain melting behavior[35]. On the other hand, it is necessary to note that we did not find any evidence of superionic transition in LiH with the same method.

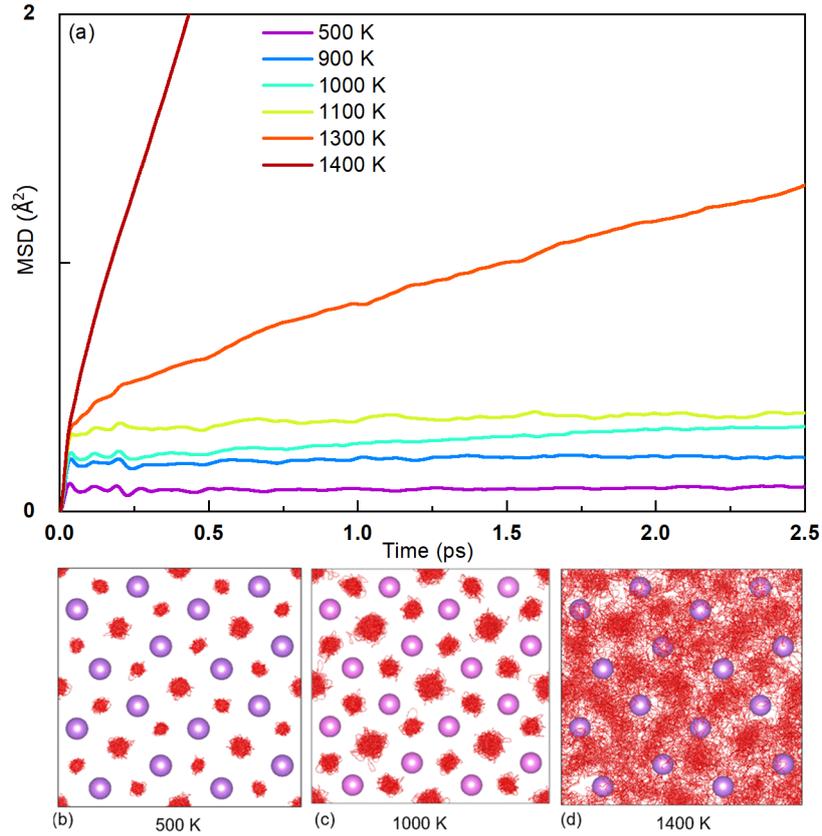

Fig. 2. (Color online) (a) Mean squared displacement of H atoms at different temperatures. (b,c,d) Trajectories of H atoms(red lines) at 500,1000 and 1400 K, respectively, in which lithium is drawn as purple balls. The atomic volume at the corresponding temperature were listed in Table.S3.

### 3.2. Melting point prediction

If temperature is increased further, the lithium sublattice also becomes unstable, and the entire system will melt and enter the liquid state. Nevertheless, in molecular dynamics simulations, superheated solid states are often encountered, mainly due to the first-order transition nature of melting, as well as the limited size of the supercell and the total simulation duration that can be accessed nowadays. In this work, Z method was employed to mitigate this problem[32][33][34]. The simulation covers specific volume ranging from 2.162 to 4.334 Å$^3$ per atom, which corresponds to a pressure of approximately from 50 to 300 GPa. The variation of $P$ and $T$ along the given isochoric path calculated by Z-method are shown in Fig.3. At the specific volume of 4.334 Å$^3$/atom, LiH$_2$ melts at 50 GPa and 1500 K. With the density increases, the melting temperature is monotonously increased to 2074 K at 300 GPa. The zigzag shape in the T-P curve along the isochoric path is evident, which demonstrates the spontaneously collapse of the lattice in the superheated regime and the subsequent relaxing to the liquid state, a feature of first-order phase transition.

The shaded area in Fig. 3 illustrates the pressure-temperature region in which the superionic state of $LiH_2$ exists. This premelting delocalization of H-sublattice could have impact on the properties of this material. The superionic transition occurs when the free energy of an atom at the lattice site becomes high enough to prefer an interstitial position, facilitated by the potential energy landscape of the liquid state[36]. The mobility of each type of atoms can be described by the MSD and the associated self-diffusion coefficient, which is defined as

$$D = \frac{1}{6N} \lim_{t \to \infty} \frac{d}{dt} \sum_{i=1}^{N} \left\langle \left| r_i(t+t_0) - r_i(t_0) \right|^2 \right\rangle \quad (1)$$

where $r_i(t)$ represents the coordinates of the particle $i$ at time t, N is the total number of this type of ions, and $\langle ... \rangle$ indicates the ensemble average.

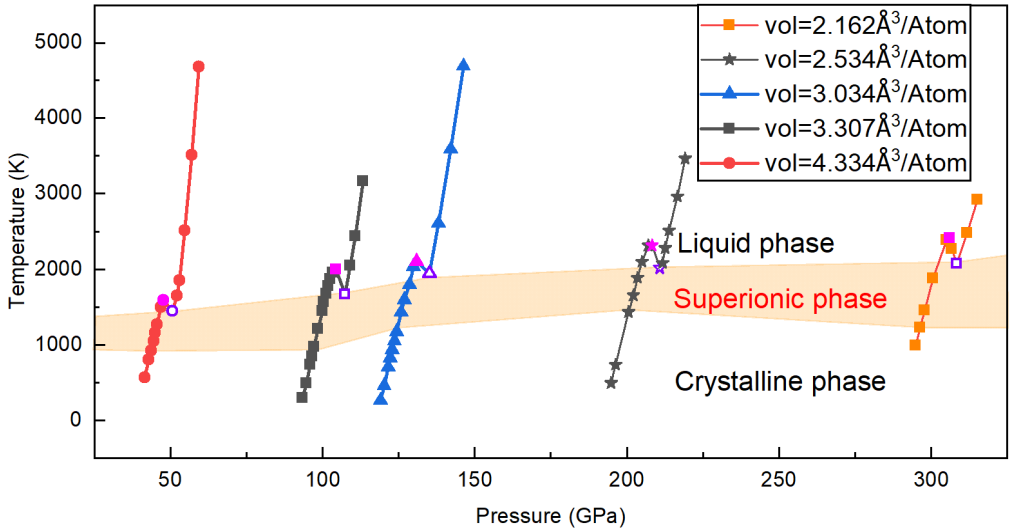

Fig. 3. (Color online) The melting points of $LiH_2$ obtained using the AIMD and Z method. The magenta filled symbols indicate the collapse point of the superheated lithium sublattice, and the open violet symbols denote the melting point.

The calculated diffusion coefficient of both hydrogen and lithium at a pressure about 130 GPa are shown in Fig.4(a). It is apparent that the increase in diffusivity of Li is delayed to higher temperature compared to that of H, suggesting the superionic nature of the phase. Furthermore, the diffusion coefficient of H does not vary linearly with pressure and temperature. This suggests that the superionic transitions are not done all at once, and there is an intermediate state between a pure solid and a liquid. During the early stage of the superionic transition, a few H atoms leave their lattice positions and begin hopping on the H sublattice via interstitial sites. However, the majority of H atoms remain in their original lattice positions. When the temperature increases further, the H sublattice eventually melts, leading to a significant jump in the diffusion coefficient of H at 1427 K, as shown in Fig. S3 of the Supplementary Material. This occurs prior to the complete melting of $LiH_2$, which is characterized by a drastic increase in the diffusivity of Li atoms at 2230 K, as illustrated in Fig. 4(a). These changes are also accompanied by variations in the P-T curve, as demonstrated in Fig. 4(b).

To explore the self-diffusion mechanism, we attempted to fit the AIMD estimated diffusivity of both H and Li before liquefaction to the following expression,

$$D = D_\infty e^{\frac{-E_a}{k_B T}} \qquad (2)$$

In Eq.(2), the parameter $E_a$ indicates the activation energy of particle hopping in a solid, $D$ represents the self-diffusion coefficient, $D_\infty$ is a constant corresponding to the asymptotic diffusivity when temperature $T \to \infty$, and $k_B$ is the Boltzmann constant. The diffusivity of H cannot be described by a single model. At the initial stage of superionicity, its D can be described by Eq.(2), in which $D_\infty = 5.9 \times 10^{-2}$ $cm^2/s$, is the asymptotic diffusivity when temperature $T \to \infty$, and the energy barrier $E_a = 1.41$ eV. When approach melting, it is described by $D = D_{T_0}(\frac{T}{T_0})^\alpha$, in which $D_{T_0} = 1.46 \times 10^{-5}$ $cm^2/s$, is the diffusivity when $T = T_0$, and $\alpha = 31.32$, the reference temperature is set to $T_0 = 2000K$. It is a mechanism of diffusion governed by particle collision, and is typical for normal fluid. In this stage, the diffusion process is significantly accelerated and primarily governed by temperature. The fitted curves are displayed in Fig.S3 of Supplementary Material.

For Li, it is well described by the thermal activation model of by Eq.(2), with $D_\infty = 1.29 \times 10^{17}$ cm$^2$/s, and the energy barrier $E_a = 9.3$ eV. From this we find that at the early stage of superionic transition, both the diffusivity of H and Li are controlled by the hopping mechanism. But the energy barrier for Li is about 6.6 times greater than that of H atoms.

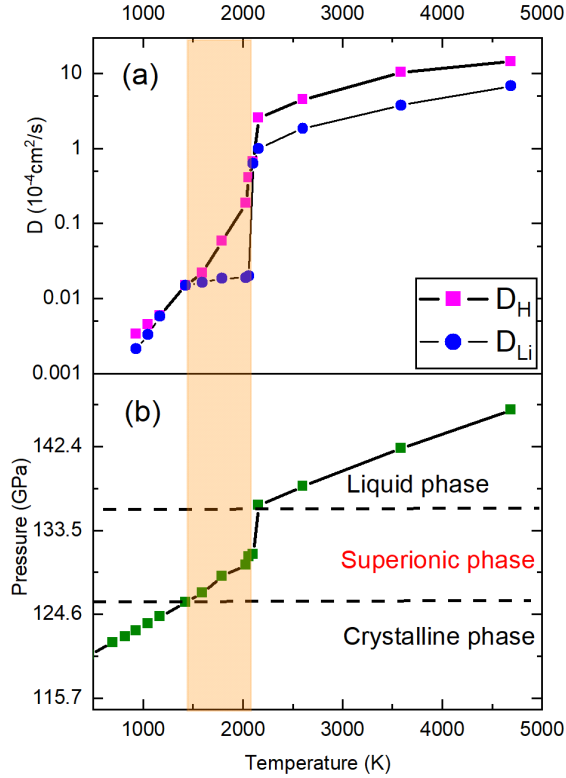

Fig. 4. (Color online) (a) Variation of the estimated diffusivity of H and Li atoms with increased temperature along the isochoric path with specific volume of 3.034 Å$^3$/atom. The shaded area indicates the superionic region. (b) The corresponding P-T curve of the same isochoric path as (a), in which the crystalline, superionic, and liquid phase are marked, respectively.

### 3.3. The change of H$_2$ fraction

In H-rich compounds there are usually quasi-molecular H$_2$ units[36], which is expected to dissociate at high pressure and temperature, much like what happens in pure hydrogen[37][38]. To investigate this scenario, we calculated the change of H$_2$ fraction when across the superionic transition and melting. The analysis method is the same as that reported in Ref.[37].

The variations of $\rho_{H_2}$ with pressure along the given isochores are plotted in Fig.5. The effect of compression is minor. The slope of these isochoric curves i.e., $\left(\dfrac{\partial \rho_{H_2}}{\partial p}\right)_V$ is almost a constant, as indicated in Fig.5. On the other hand, we found that the driven factor for $H_2$ dissociation is the temperature, as demonstrated in Fig.6.

It is necessary to point out that for most densities we studied, $\rho_{H_2}$ decreases almost monotonically with temperature, except the low pressure one with an atomic volume of 5.294 Å$^3$/atom, for which $\rho_{H_2}$ firstly increases with temperature to a maximum, and then decreases. The $H_2$ fraction reaches the highest when the superionic state appears, indicating that mobility of H allows them to recombine into $H_2$ units. We also found that the rapid decrease of $\rho_{H_2}$ does not correlate to the melting process of LiH$_2$.

Since both linear or compact H$_3$ clusters were predicted in the ground state of H-rich compounds[5][39] and at around the dissociation region of pure H[37][40][41], we also estimate the fraction and lifetime of H$_3$ units. According to Fig.S4 the lifetime of H$_3$ cluster is very short, and increasing temperature will increase its fraction, as shown in Fig.S5. However, when beyond 1647 K, its lifetime starts to decrease with increasing temperature, indicating that these H$_3$ clusters are just accident encounter under these conditions, very similar to what happens in pure H[37].

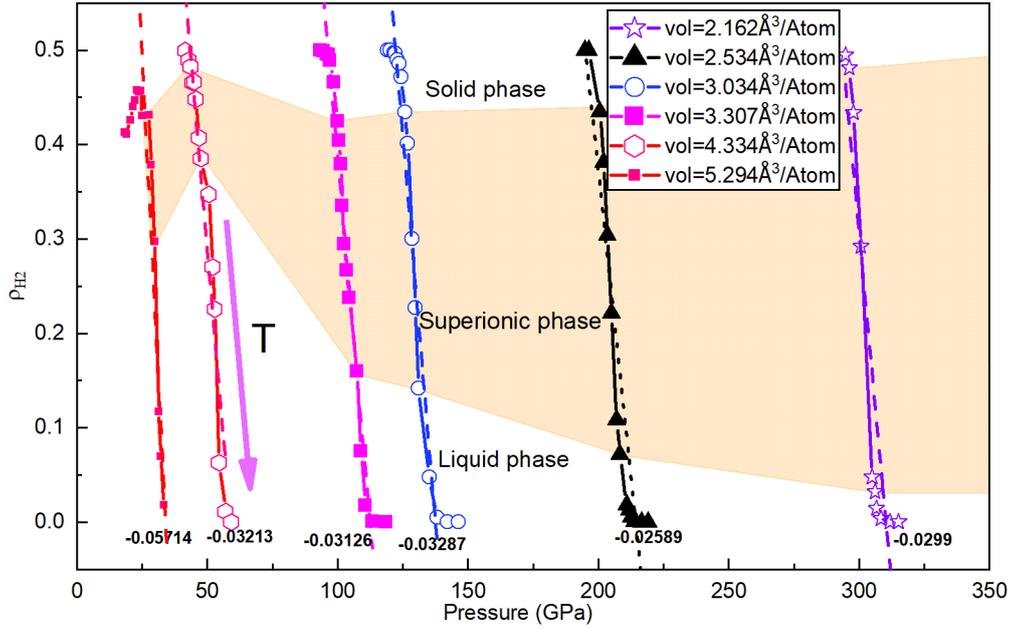

Fig. 5. (Color online) Variation of the fraction of H$_2$ units in LiH$_2$ with increased temperature along given isochoric paths, in which the corresponding region of solid, superionic state and liquid are marked out, respectively. The purple arrow shows the direction of the temperature increment. The dotted lines denote the slope of the curves $\left(\frac{\partial \rho_{H_2}}{\partial p}\right)_V$, as the number next to the line indicated.

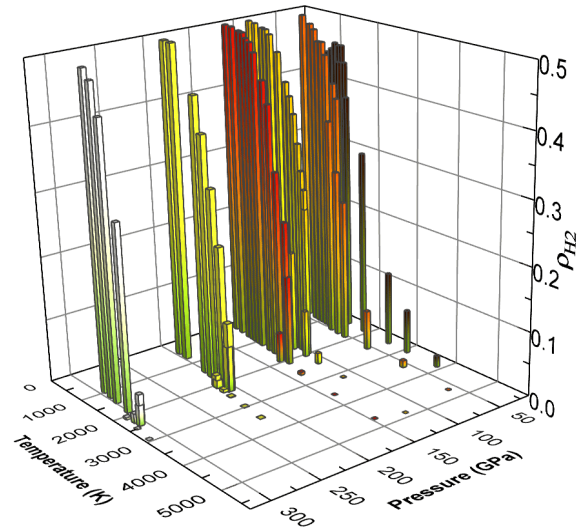

Fig.6. (Color online) Variation of $\rho_{H_2}$ with both temperature and pressure. Different color represents difference atomic volume.

### 3.4. Elastic properties

By solving the stress-strain relation that sampled from AIMD simulations[31], we obtained the independent elastic constants $C_{11}$, $C_{12}$, $C_{13}$, $C_{33}$, $C_{44}$, and $C_{66}$ of LiH$_2$, which are shown in Fig. 7(a). The mechanical stability is confirmed by using the Born stability criterion[42]

$$C_{11} > |C_{12}|, 2C_{13}^2 < C_{33}(C_{11}+C_{12}), C_{44} > 0, C_{66} > 0 \quad (3)$$

We found that except $C_{13}$, almost all elastic constants are softened when entering the superionic state. Furthermore, using these single-crystalline $C_{ij}$, we also obtained the polycrystalline modulus and the corresponding sound velocities, as listed in Table I and showing in Fig.7(b), respectively. Our results provide direct numerical evidence of the strong pre-melting softening in LiH$_2$ due to the delocalization of H sublattice in superionic state.

Table I: Calculated bulk modulus(B), shear modulus(G), Young's modulus(E), and Poisson's ratio($\nu$) of LiH$_2$ at 50 GPa.

| T(K) | B (GPa) | G (GPa) | E (GPa) | $\nu$ |
|---|---|---|---|---|
| 100 | 163 | 69 | 181 | 0.32 |
| 300 | 152 | 61 | 162 | 0.32 |
| 500 | 152 | 60 | 160 | 0.33 |
| 700 | 148 | 55 | 148 | 0.33 |
| 900 | 146 | 52 | 140 | 0.34 |
| 1100 | 131 | 31 | 85 | 0.39 |
| 1200 | 119 | 35 | 94 | 0.37 |
| 1300 | 100 | 16 | 45 | 0.43 |
| 1400 | 125 | 22 | 63 | 0.42 |

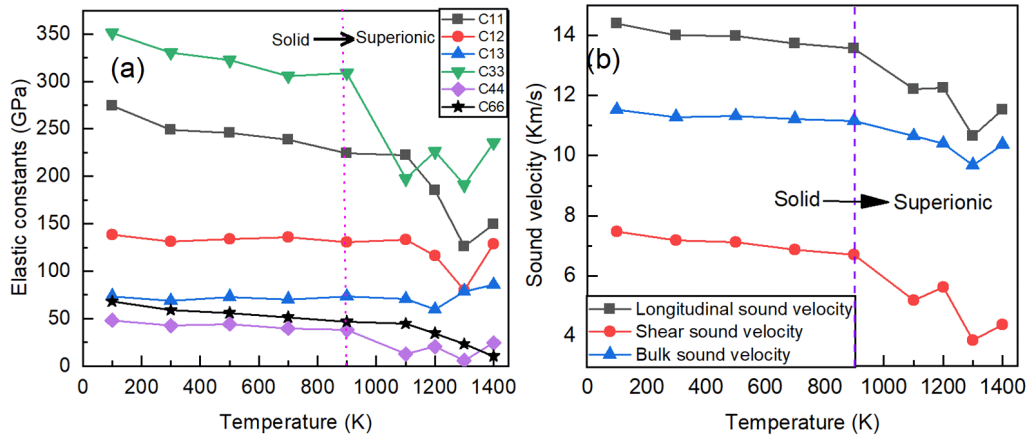

Fig. 7. (Color online) (a) The calculated monocrystalline elastic constants and (b) polycrystalline sound velocities of $LiH_2$ as a function of temperature at 50 GPa.

## 4. Conclusion

The structure and properties of $LiH_2$ at high pressures and temperatures were investigated with AIMD at the level of vdW-DF exchange correlation functional approximation of DFT. The melting temperature of $LiH_2$ up to 300 GPa was determined. A peculiar superionic transition before melting was predicted, which has an 1D chain melting in the H sublattice at the early stage of the superionic transition. A change in the diffusion mechanism within the superionic state is also identified. The analysis of the variation of $H_2$ and $H_3$ units reveals that this hydride bears some similarities with the pure H in the vicinity of dissociation. Furthermore, a strong softening in elastic constants and sound velocity of $LiH_2$ caused by the premelting superionic transition was predicted. This phenomenon has broad implications beyond $LiH_2$ and may be relevant to other H-rich compounds. We also note that there is no superionic state in LiH. This anomaly in $LiH_2$ may have impact on the quasi-isentropic compression. Consequently, it could have far-reaching implications for optimizing the compression path of ICF if materials like lithium tritium-deuteride(LiTD) an isotope of $LiH_2$ are utilized as the nuclear fusion fuels in the future. Our results also provide insightful understanding about the high pressure and high temperature behavior of H-rich compounds that are promising for applications in hydrogen storage and near room temperature superconductor.

**Acknowledgment**

This work was supported by National Key R&D Program of China under Grant No. 2021YFB3802300, the National Natural Science Foundation of China under Grant No. 11672274 and the NSAF under Grant No. U1730248. Part of the computation was performed using the supercomputer at the Center for Computational Materials Science (CCMS) of the Institute for Materials Research (IMR) at Tohoku University, Japan.

# Supplementary material for "Prediction of superionic state in LiH$_2$ at conditions enroute to nuclear fusion"

Fude Li(李福德)[1], Hao Wang(王豪)[1], Jinlong. Li(李津龙)[1], Hua Y. Geng(耿华运)[1,2][1]

[1]*National Key Laboratory of Shock Wave and Detonation Physics, Institute of Fluid Physics, CAEP; P.O.Box 919-102 Mianyang, Sichuan, 621900, China*

[2]*HEDPS, Center for Applied Physics and Technology, and College of Engineering, Peking University, Beijing 100871, China*

Table S1: Structure of LiH$_2$ with a space group of P4/mbm at 150 GPa.

| Lattice parameters | Atomic coordinates |
| --- | --- |
| a = b = 4.115 Å | H1(4e) (0.0000, 0.0000, 0.3135) |
| c = 1.967 Å | H2(4g) (0.8509, 0.6490, 0.0000) |
| α = β = γ = 90.000º | Li(4h) (0.8453, 0.3453, 0.5000) |

Table S2: Superionic transition temperature of LiH$_2$ at given densities.

| Superionic temperature (K) | Atomic volume (Å$^3$/atom) |
| --- | --- |
| 1026 | 5.294 |
| 900 | 4.334 |
| 967 | 3.307 |
| 1167 | 3.034 |
| 1430 | 2.534 |

Table S3: The atomic volume of LiH$_2$ varies at 50 GPa and different temperatures.

| Temperature (K) | Atomic volume (Å$^3$/atom) |
| --- | --- |
| 500 | 4.157 |
| 900 | 4.263 |
| 1000 | 4.298 |

[1] Email: s102genghy@caep.cn

| 1100 | 4.399 |
| 1300 | 4.412 |
| 1400 | 4.635 |

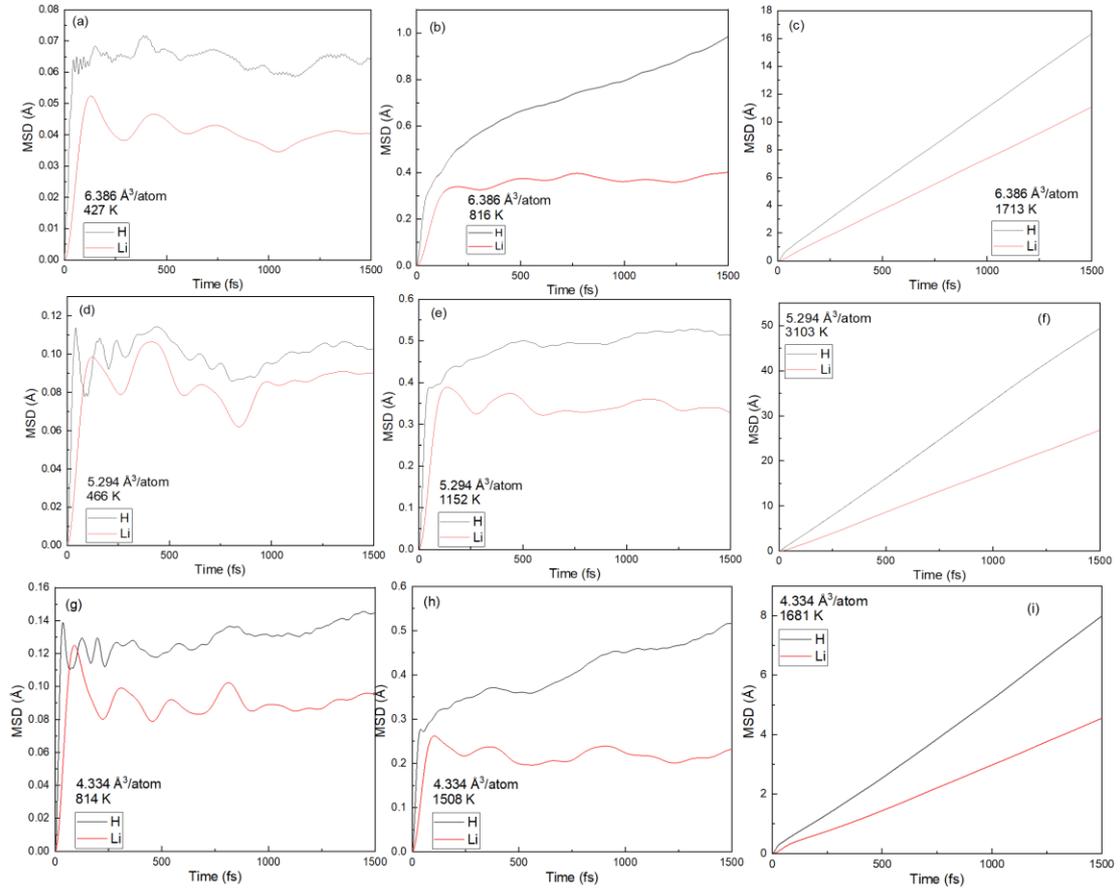

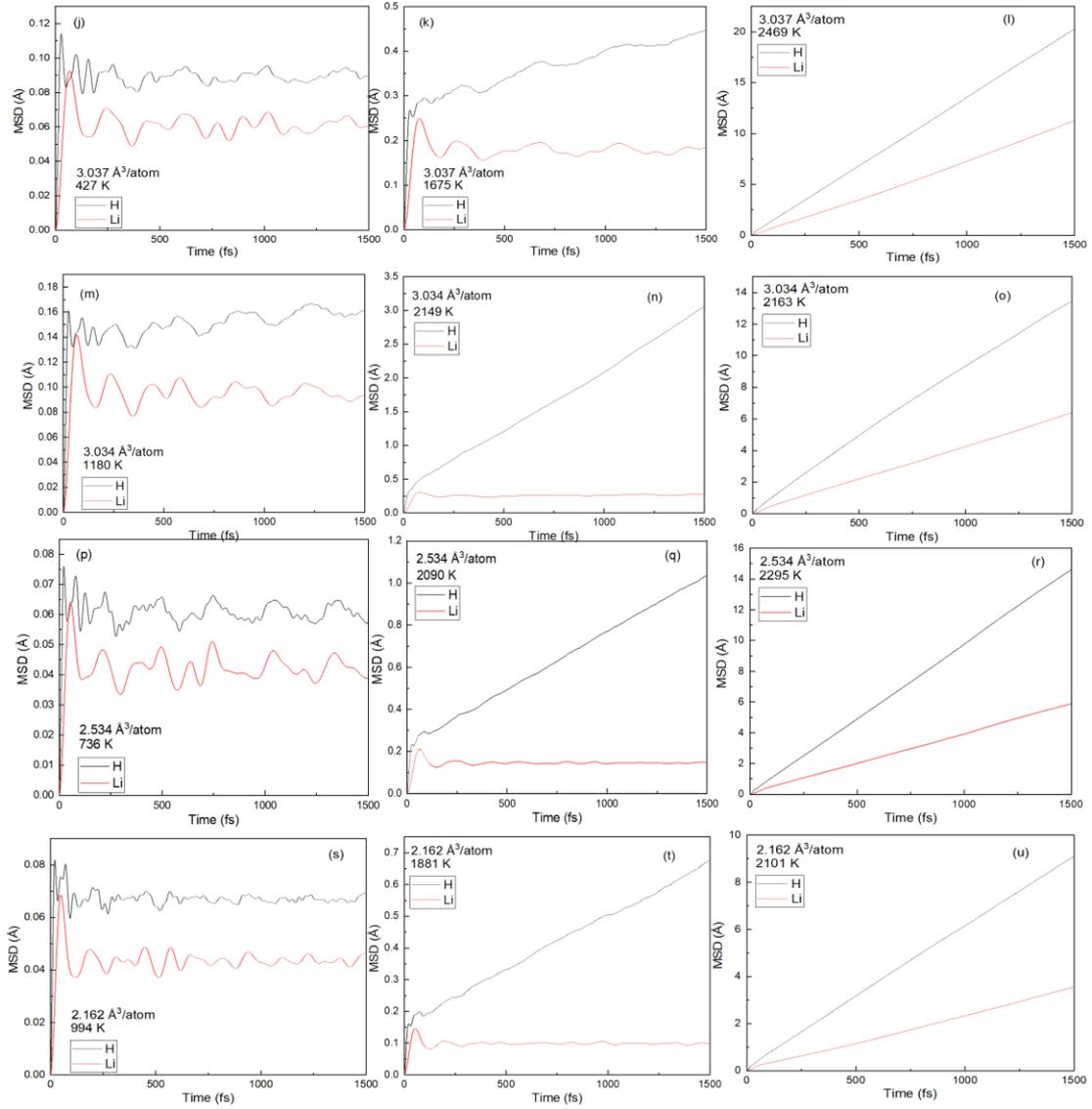

Figure. S1. (Color online) Evidence of superionic phase transition at different atomic volumes.

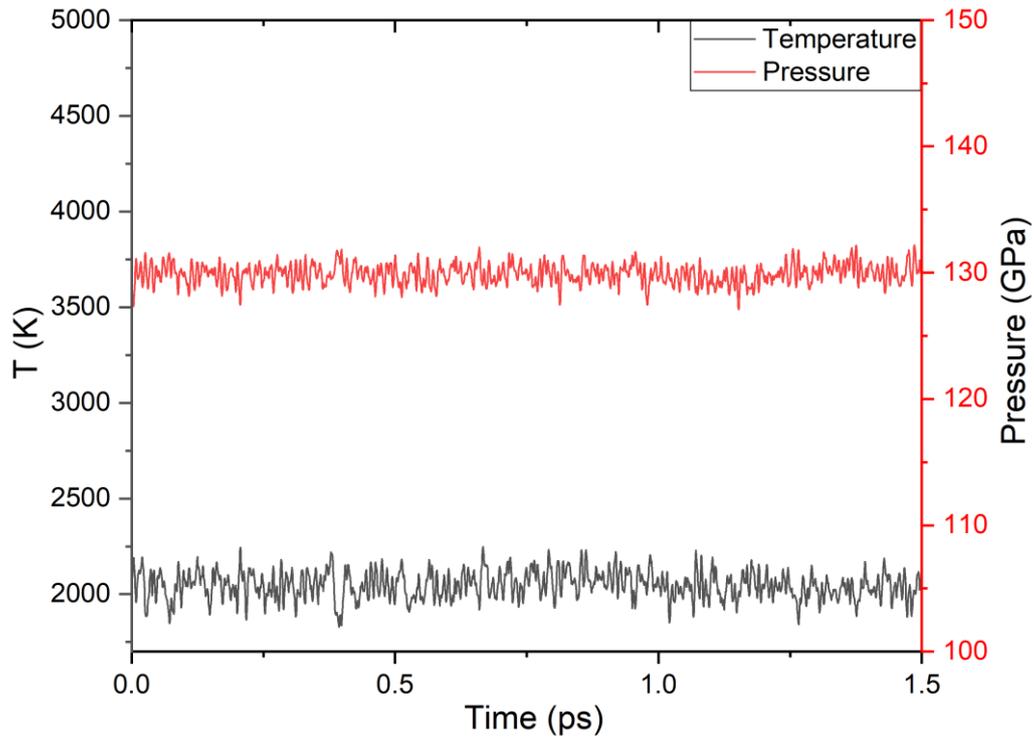

Figure. S2. (Color online) The evolution of temperature and pressure in AIMD with respect to the simulation time for a density with specific volume of 3.034 Å$^3$/atom.

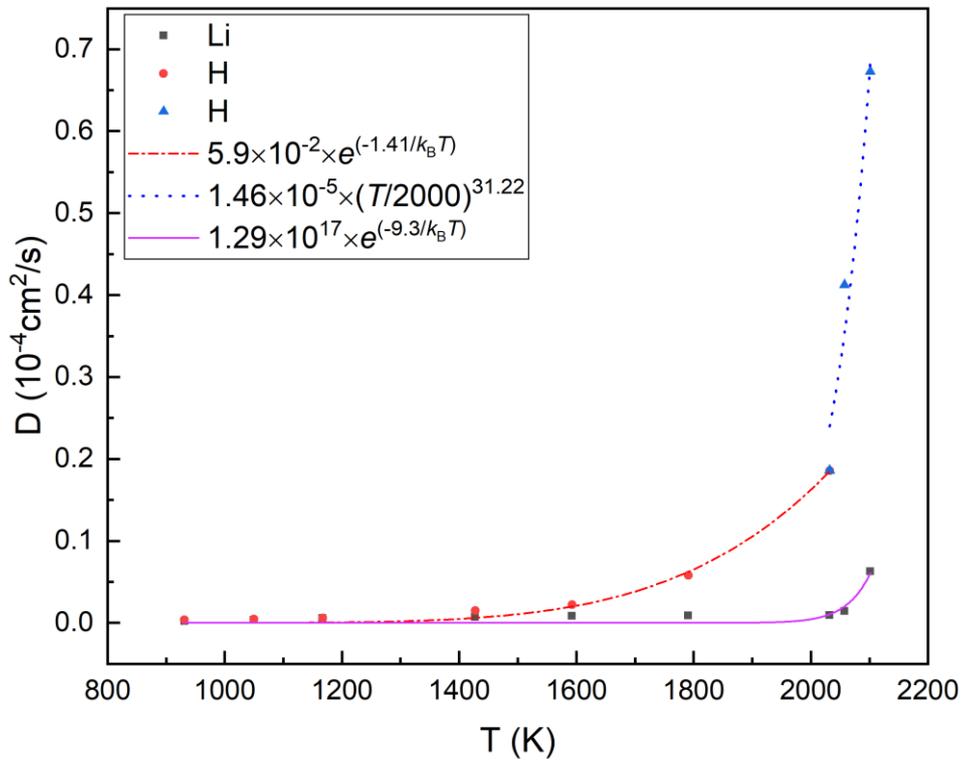

Figure. S3. (Color online) The AIMD calculated diffusivity (solid symbols) of both H

and Li atoms compared with the model fitted curves before liquefaction in LiH$_2$ with a specific volume of 3.034 Å$^3$/atom.

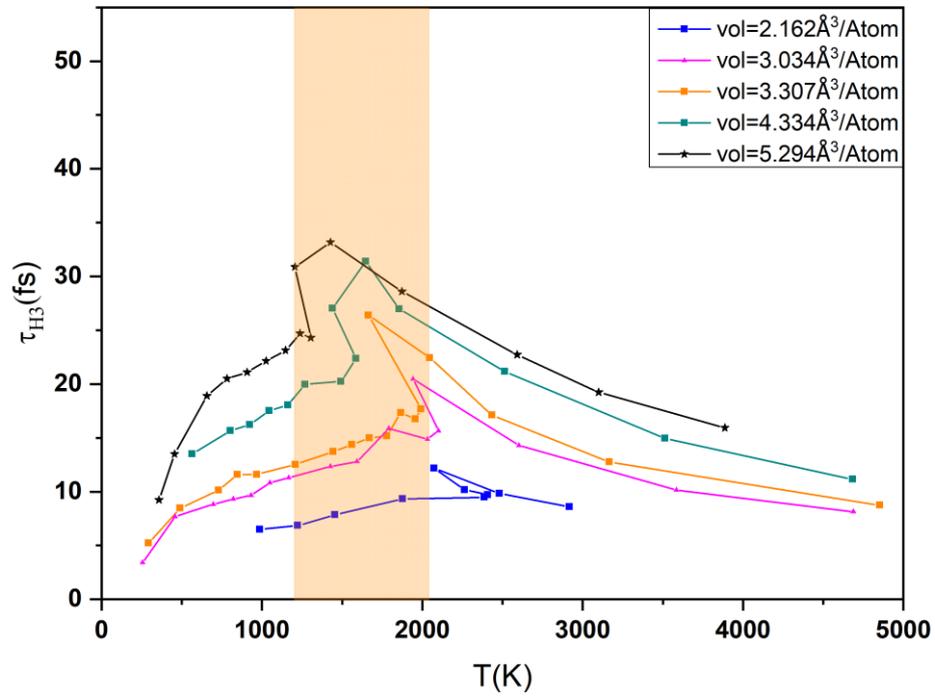

Figure. S4. (Color online) The AIMD estimated lifetime of H$_3$ units at various temperatures and densities. The shaded area indicates the temperature range where H$_3$ units reach the longest lifetime.

Chinese Physics B **32**, 106103 (2023)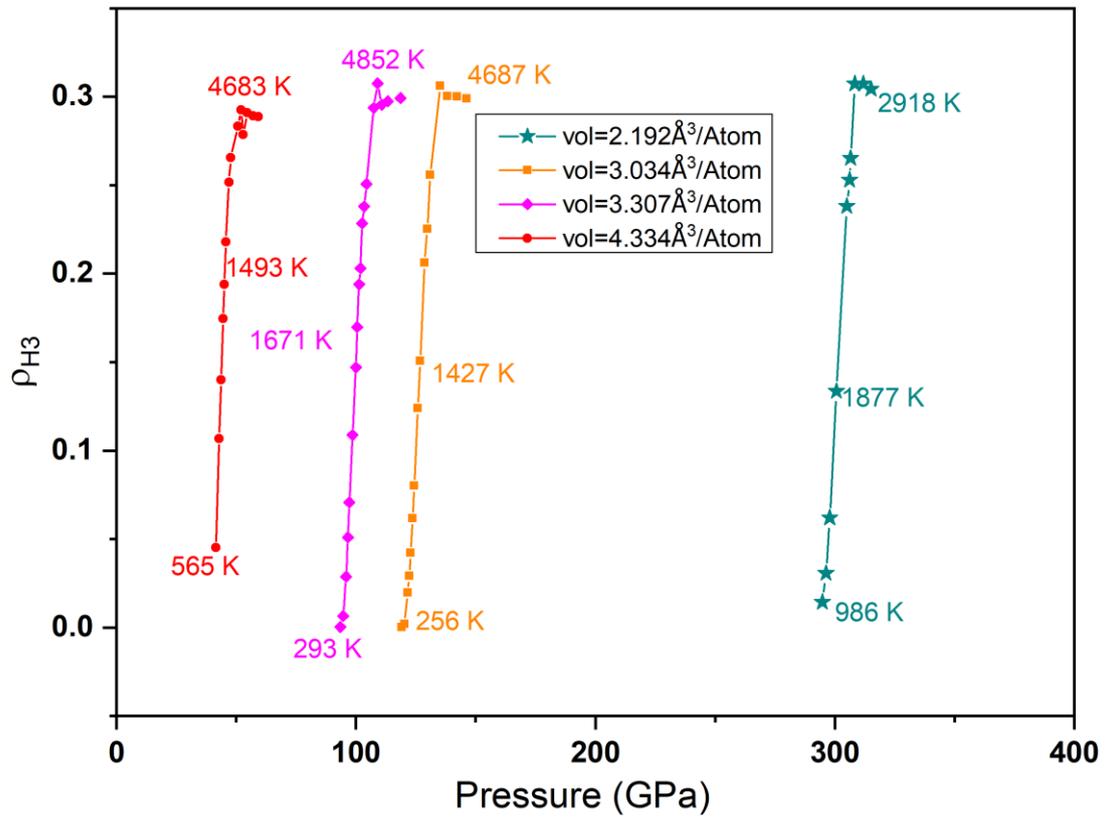

Figure. S5. (Color online) The change in H$_3$ fraction with temperature along an isochoric path at various densities. The temperatures of some specific points along the path are also labeled. It is important to note that the pressure variation along each curve is due to the thermal pressure with increasing temperatures.

23